\documentclass[12pt,preprint]{aastex}

\def\gs{\mathrel{\raise0.35ex\hbox{$\scriptstyle >$}\kern-0.6em
\lower0.40ex\hbox{{$\scriptstyle \sim$}}}}
\def\ls{\mathrel{\raise0.01935ex\hbox{$\scriptstyle <$}\kern-0.6em
\lower0.40ex\hbox{{$\scriptstyle \sim$}}}}

\input{psfig.sty}

\shorttitle{{\em Spitzer} observations of MAMBO galaxies}
\shortauthors{Ivison et al.}

\begin{document}

\title{{\it Spitzer} observations of MAMBO galaxies: weeding out
\newline active nuclei in starbursting proto-ellipticals}

\author{R.\,J.\ Ivison,\altaffilmark{1,2}
T.\ R.\ Greve,\altaffilmark{2}
S.\ Serjeant,\altaffilmark{3}
F.\ Bertoldi,\altaffilmark{4}
E.\ Egami,\altaffilmark{5}
A.\ M.\ J.\ Mortier,\altaffilmark{3}\newline
A.\ Alonso-Herrero,\altaffilmark{5}
P.\ Barmby,\altaffilmark{6}
L.\ Bei,\altaffilmark{5}
H.\ Dole,\altaffilmark{5}
C.\ W.\ Engelbracht,\altaffilmark{5}
G.\ G.\ Fazio,\altaffilmark{6}\newline
D.\ T.\ Frayer,\altaffilmark{7}
K.\ D.\ Gordon,\altaffilmark{5}
D.\ C.\ Hines,\altaffilmark{5}
J.-S.\ Huang,\altaffilmark{6}
E.\ Le Floc'h,\altaffilmark{5}
K.\ A.\ Misselt,\altaffilmark{5}\newline
S.\ Miyazaki,\altaffilmark{8}
J.\ E.\ Morrison,\altaffilmark{5}
C.\ Papovich,\altaffilmark{5}
P.\ G.\ P\'erez-Gonz\'alez,\altaffilmark{5}
M.\ J.\ Rieke,\altaffilmark{5}\newline
G.\ H.\ Rieke,\altaffilmark{5}
J.\ Rigby,\altaffilmark{5}
D.\ Rigopoulou,\altaffilmark{9}
I.\ Smail,\altaffilmark{10}
G.\ Wilson\altaffilmark{7} and
S.\ P.\ Willner\altaffilmark{6}}

\altaffiltext{1}{Astronomy Technology Centre, Royal Observatory,
Blackford Hill, Edinburgh EH9 3HJ UK}
\altaffiltext{2}{Institute for Astronomy, University of Edinburgh,
Blackford Hill, Edinburgh EH9 3HJ UK}
\altaffiltext{3}{Centre for Astrophysics \& Planetary Science, School
of Physical Sciences, University of Kent, Canterbury, Kent CT2 7NR UK}
\altaffiltext{4}{Max-Planck-Institut f\"ur Radioastronomie, Auf dem
H\"ugel 69, 53121 Bonn, Germany}
\altaffiltext{5}{Steward Observatory, University of Arizona, 933 N
Cherry Avenue, Tuscon, AZ 85721}
\altaffiltext{6}{Harvard-Smithsonian Center for Astrophysics, 60
Garden Street, Cambridge, MA 02138}
\altaffiltext{7}{Spitzer Science Center, California Institute of Technology,
Mail Code 220-6, 1200 East California Boulevard, Pasadena, CA 91125}
\altaffiltext{8}{Subaru Telescope, National Astronomical Observatory of Japan,
650 North Aohoku Place, Hilo, HI 96720}
\altaffiltext{9}{Astrophysics, Denys Wilson Building, Keble Road,
Oxford OX1 3RH, UK}
\altaffiltext{10}{Institute for Computational Cosmology, University of Durham,
South Road, Durham DH1 3LE, UK}

\setcounter{footnote}{11}

\begin{abstract}
We present {\em Spitzer} observations in five wavebands between 3.6 and
24\,$\mu$m of an unbiased sample of nine luminous, dusty galaxies selected at
1200\,$\mu$m by the MAMBO camera on the IRAM 30-m telescope, a population akin
to the well-known submillimeter (submm) or `SCUBA' galaxies (hereafter
SMGs). Owing to the coarse resolution of submm/mm instrumentation, SMGs have
traditionally been difficult to identify at other wavelengths. We compare our
multi-wavelength catalogs to show that the overlap between 24 and 1200\,$\mu$m
must be close to complete at these flux levels.  We find that all (4/4) of the
most secure $\ge$4$\sigma$ SMGs have robust $\ge$4$\sigma$ counterparts at
1.4\,GHz, while the fraction drops to 7/9 using all $\ge$3$\sigma$ SMGs.  We
show that combining mid-infrared (mid-IR) and marginal ($\ge$3$\sigma$) radio
detections provides plausible identifications in the remaining cases, enabling
us to identify the complete sample.  Accretion onto an obscured central engine
is betrayed by the shape of the mid-IR continuum emission for several sources,
confirming {\em Spitzer}'s potential to weed out active galaxies. We
demonstrate the power of a $S_{\rm 24\mu m}/S_{\rm 8\mu m}$ versus $S_{\rm 8\mu
m}/S_{\rm 4.5\mu m}$ color-color plot as a diagnostic for this
purpose. However, we conclude that the majority ($\sim$75\%) of SMGs have
rest-frame mid-/far-IR spectral energy distributions (SEDs) commensurate with
obscured starbursts.  Sensitive 24-$\mu$m observations are clearly a useful
route to identify and characterize reliable counterparts to high-redshift
far-IR-bright galaxies, complementing what is possible via deep radio imaging.
\end{abstract}


\section{Introduction}

Major progress has been made in our understanding of galaxy formation
and evolution since the discovery of a significant population of
submm/mm-bright, dusty galaxies. The first handful of galaxies,
discovered behind lensing clusters \citep{sib97}, and in the Hubble
Deep Field \citep{hug98}, revealed the diversity of the population.
SCUBA \citep{hol99}, the innovative camera with which SMGs were first
seen, was
commissioned on the James Clerk Maxwell Telescope during 1996--7. Since
then, the 117-bolometer, 1200-$\mu$m MAMBO camera \citep{ber00} has
been installed on the IRAM 30-m telescope, Spain. Surveys at
1200\,$\mu$m are sensitive to the same dusty galaxies as SCUBA, yet
with a selection function that potentially stretches to higher
redshifts and is sensitive to lower dust temperatures
\citep{eal03,bla04b}.

A moderate fraction of SMGs can be pinpointed by their $\mu$Jy-level
emission at 1.4\,GHz (Ivison et al.\ 2002), allowing us to identify
their counterparts in other wavebands.  However, up to 30\% of the
population remain undetected in the radio waveband in even the deepest
maps and the properties of these galaxies remain a
mystery.  Those SMGs which can be localised through their radio
emission are usually found in the optical/IR wavebands to be faint,
morphologically complex systems, often comprising red galaxies with
bluer companions, as expected for a distant, dust-reddened, interacting
star-forming population \citep{ivi02,sma02,web03}.  Recent
advances in their study include the measurement of the redshift
distribution of radio-detected SMGs \citep{cha03}, the
resolution of most of the submm background by exploiting gravitational
lensing \citep{bla99,cow02}, the identification of X-ray emission from
a significant fraction of the population \citep{ale03}, the detection
--- via broad CO lines --- of collosal quantities
($\sim$10$^{11}$\,M$_{\odot}$) of molecular gas
\citep{fra98,ner03,gre04a}, and the first indications of the strong
clustering expected for such massive galaxies
\citep{bla04a}.  These discoveries underline the
importance of studying the far-IR luminous SMGs for our understanding
the formation of massive galaxies at high redshifts.

For local IR-luminous galaxies, mid-IR data from the {\em Infrared
Space Observatory (ISO)} were used to develop a number of useful
diagnostics to differentiate between those powered predominantly by
starbursts and by active galactic nuclei (AGN) \citep{gen00}. {\em
ISO} was sensitive to only the most luminous SMGs, so efforts
to assess the preponderance of AGN in SMGs have so far been reliant on
UV/optical spectroscopy, radio observations, and the aforementioned
X-ray imaging, often with ambiguous results. The fraction of distant,
far-IR-luminous galaxies with an energetically dominant AGN remains
poorly determined, although it is expected to be small: even SMGs with
unambiguous AGN characteristics \citep{ivi98} are thought to have
a major starburst contribution to their bolometric luminosities
\citep{fra98,gen03}. This finding has been echoed in studies of
distant quasars and radio galaxies: far-IR-luminous examples are found
to be gas-rich and their submm emission is often resolved, as expected
if their power originates in large part from stars
\citep{omo96,pap00,ste03}.

In this Letter, we use mid-IR continuum imaging of nine MAMBO-selected
SMGs in the Lockman Hole to investigate the use of mid-IR data to
identify the counterparts at other wavelengths of this population and
demonstrate the power of mid-IR data for determining the likely power
source (AGN or starburst) in these galaxies.

\section{MAMBO galaxies and the diagnostic potential of {\em Spitzer}}

\citet{gre04b} reported the results of a deep unbiased MAMBO survey of the
Lockman Hole, a region mapped earlier by
SCUBA for the `8-mJy Survey' \citep{sco02}.  This region was selected for
reasons that included its low Galactic cirrus emission and the availability of
high-quality complementary data at X-ray, optical/IR and radio wavelengths.
Not surprisingly, the same region was selected for some of the earliest imaging
with {\em Spitzer} \citep{wer04}. The data exploited herein were obtained using
IRAC \citep{faz04} and MIPS \citep{rie04}, reaching noise levels of $\sigma$ =
0.8, 0.8, 2.8, 1.7 and 30\,$\mu$Jy at 3.6, 4.5, 5.8, 8.0 and 24\,$\mu$m with
absolute calibration accurate to $\pm10$\%. These compare with 5-$\sigma$
confusion limits (20 beams per source) of around 1.0, 1.2 and 56\,$\mu$Jy at
3.6, 8.0 and 24\,$\mu$m \citep{vai01,dol04}. The images cover a $\rm 5' \times
5'$ region, centered near the SMG, LE\,850.01 (Table~1). Data
were reduced using standard {\em
Spitzer} data analysis tools, as described by \citet{lef04} and \citet{hua04}.

Our analysis also makes use of deep optical imaging from SuprimeCam on the
Subaru 8-m telescope ($R, \rm 3\sigma\sim 27.0$, Vega, 2$''$-radius aperture),
{\em XMM-Newton} X-ray data ($\rm 3\sigma\sim 5\times 10^{-16}\, ergs\, s^{-1}
\,cm^{-2}$, 0.5--10\,keV) and a Very Large Array radio image ($\rm 3\sigma\sim
15\, \mu Jy\, beam^{-1}$, 1.4\,GHz) \citep{ivi02}.

The \citet{gre04b} 1200-$\mu$m sample includes nine $\ge$3$\sigma$ MAMBO
sources within the region covered by the IRAC/MIPS
observations.  Four are detected at $\ge$\,4$\sigma$,
all of which should be secure. The other five are $\ge$\,3-$\sigma$ detections,
of which up to two may be the result of noise fluctuations, Eddington bias
\citep{sco02} or confusion. \citet{gre04b} performed
Monte Carlo simulations to determine the reliability, completeness
and positional accuracy of their source catalog.  We list the flux
densities, significance and positions for all nine sources in Table~1 and note
that the positions are accurate to $\pm$4$''$ (95\% confidence).

Fig.~1 shows the SEDs of typical local starburst- and AGN-dominated IR-luminous
galaxies and several SMGs with well-sampled UV--radio spectra.  It is worth
reviewing the emission mechanisms operating across the observable wavelength
range: where our adopted sample was selected, at 1200\,$\mu$m (rest-frame
far-IR), we are sensitive to cold dust ($\sim$40\,{\sc k}) created in copious
quantities by supernovae (SNe; \citet{dun03}), which re-radiates energy
absorbed in the UV from hot, young stars. In the radio, at 1.4\,GHz, we are
again sensitive to SNe (and hence to recent star formation), via synchrotron
radiation from relativistic electrons. Contamination via radio-loud AGN is
possible and, at $\mu$Jy flux levels, is virtually impossible to distinguish
from pure SNe-related emission.  We are also sensitive to AGN via X-ray
emission from accretion disks and their associated coronae, although
determining the origin of the X-rays is not always trivial and heavily obscured
($N({\rm H})>10^{24.5}$\,cm$^{-2}$, Compton-thick) AGN can evade detection
altogether \citep{ale03}.  At the shortest IR wavelengths accessible to {\em
Spitzer} (3.6--8.0\,$\mu$m), we are provided --- via photospheric emission from
stars --- with a relatively unobscured measure of stellar luminosity, possibly
even of stellar mass in more evolved systems out to $z>2$. At 24\,$\mu$m
(rest-frame $\sim$7\,$\mu$m for a galaxy at $z\sim 2$--3) we are sensitive to
emission from $\sim$500-{\sc k} dust in the circumnuclear torus of AGN, and to
the warmest dust in starbursts.

Do our {\em Spitzer} observations provide
sufficient sensitivity to detect dust-enshrouded starbursts and AGN in the
distant Universe?  In Fig.~1 we illustrate the SEDs of a selection of
well-studied dusty, luminous galaxies and models (both AGN and starbursts).
These demonstrate the range of rest-frame mid-IR luminosities for these systems
and we select Arp\,220 and Mrk\,231 to represent extremes of the the starburst-
and AGN-dominated sub-classes. Fig.~2 shows that even with its steep mid-IR
spectrum ($\rm \alpha= -2.9$, where $F_{\nu} \propto \nu^{\alpha}$), Arp\,220
would be detected at 3.6 and 24\,$\mu$m out to $z\gs \rm 1$ (see also
\citet{bla04b}). Mrk\,231, its mid-IR emission boosted an
order of magnitude higher than Arp\,220 by its AGN, would be detected out to
$z\gs \rm 3$ at 24\,$\mu$m.

The diagnostic power of the near- and mid-IR bands --- the ability to
discriminate between starbursts with and without buried, active nuclei ---
arises from the different physical regimes and spectral features probed across
the mid-IR waveband \citep{rig02}. A steeper slope between rest-frame
$\sim3$--10\,$\mu$m is apparent for starbursts than for AGN-dominated galaxies.
The latter, typified by Mrk\,231 and SMM\,J02399$-$0136 (Fig.~1), have
power-law spectra covering rest-frame $\sim$0.2--10\,$\mu$m \citep{ivi98}. In
contrast, the SED of a starburst like Arp\,220 has a flatter region between 1
and 4\,$\mu$m. Thus for MAMBO galaxies at $z\sim\rm 2$--3 (assuming the same
median redshift as for SCUBA galaxies --- \citet{cha03}), the key spectral
indices are available in the 3.6--24-$\mu$m bands covered by IRAC and
MIPS. Fig.~3, a $S_{24\mu \rm m}/S_{8\mu \rm m}$ versus $S_{8\mu \rm
m}/S_{4.5\mu \rm m}$ color--color plot, shows the tracks of Arp\,220 and
Mrk\,231 as they are moved from the local Universe to $z\gg \rm 1$, as well as
the track of a synthetic $10^{13}$-L$_{\odot}$ starburst from
\citet{lag03}.  We can see immediately that the starburst and AGN populations
are well separated, and that this plot has strong diagnostic potential: we can
define regions where starburst and active galaxies are likely to lie at $z =
1$--3 and hence use this to classify our SMGs.

\section{Mid-IR counterparts to MAMBO galaxies}

We search for radio and mid-IR counterparts to the MAMBO sources within their
nominal error circles (illustrated in Fig.~4). We find that all of the four
most secure $\ge$4$\sigma$ MAMBO sources have robust radio counterparts at
1.4\,GHz (as defined by \citet{ivi02}).  This drops slightly to 7/9 robust
radio counterparts using all $\ge$3$\sigma$ MAMBO sources, somewhat higher than
the 18/42 detection rate\footnote{\citet{dan04} relied on a 1.4-GHz image which contains
several stripe-inducing $>$60-mJy sources, with a noise level 2--3 times
higher than the radio data utilized here. Using equivalent data, our fraction
of radio-detected MAMBO sources would fall to 5/9, consistent with
\citet{dan04}.} reported by \citet{dan04}. The remaining
two have tentative radio identifications at $\ge$3$\sigma$ -- but these are not
significant enough to be reliable. If instead we had used the 24\,$\mu$m {\em
Spitzer} imaging to identify counterparts we would have identified 8/9 of the
$\ge$3$\sigma$ MAMBO galaxies with significant 24\,$\mu$m sources.
Interestingly, all but one of the seven MAMBO sources with robust radio
counterparts are detected at 24\,$\mu$m, the exception being
MM\,J105207.2+572558.  Combining the two identification schemes increases our
confidence in the identification of the proposed counterparts in both wavebands
and provides reliable identifications for {\it all} nine SMGs in this region.

Detections at 24\,$\mu$m ({\sc fwhm} $\sim$ 6$''$) halve the mm
positional uncertainties (to $\pm$2$''$, 95\% confidence), although the
accuracy remains short of that provided by radio imaging
($\pm$0.6--1.6$''$, 95\% confidence --- \citet{ivi02}). Hence the radio
remains the most useful waveband for localising the far-IR emission
given sufficiently long integrations.

\section{Weeding out AGN}

We next investigate the power of mid-IR observations to identify AGN within the
SMG population.  Only one of our nine MAMBO sources has an X-ray counterpart,
MM\,J105200.2+572425 (= LE\,850.08), previously identified by \citet{ivi02}. It
could be argued that the complexity of this immense merger --- there are at
least seven distinct optical components in Fig.~4 --- is such that it would be
surprising if at least one of the galaxies were not active! In the absence of
the X-ray detection, could we have identified which components contain buried
AGN using only the mid-IR data? We could. We list the fluxes of the three
resolved mid-IR components in this system separately in Table~1 to check
whether they have different SEDs. Fig.~1 shows the individual SEDs, arbitrarily
placed at $z=\rm 2.5$. The component near-coincident with the brightest {\em
XMM-Newton} and radio emission (`b' in Fig.~4; see also \citet{ivi02}) has a
power-law mid-IR spectrum reminiscent of Mrk\,231. Component `a' (previously
assumed to be the source of X-rays, based on a less precise centroid from {\em
ROSAT}) was identified as an AGN at $z=\rm 0.97$ by \citet{leh01}) and shows a
weak break in slope at an observed wavelength around 4$\mu$m, suggesting a
possible mix of AGN and star formation power sources.  Component `c', on the
other hand, has a strong short-wavelength upturn indicative of a starburst.
This system allows us to demonstrate the diagnostic power of Fig.~3, with
component `b' lying close to the track of Mrk\,231, well separated from the
starburst track; component `a' lies between the AGN and starburst tracks and
within $\Delta z\rm \ls 0.1$ of its predicted redshift based on the SED of
Mrk\,231; component `c' is classified as a pure starburst.  The main area of
possible confusion with this diagram is separating high-redshift starbursts
from low-redshift AGN; fortunately, as in this case, we can expect AGN at
$z\sim\rm 1$ to reveal themselves in other wavebands. On the basis of these
tests, it seems reasonable to expect that mid-IR observations with {\em
Spitzer} will be able to weed out active galaxies within the SMG population.

The mid-IR data for the other SMGs shown in Figs~1 and 3 display a range
of characteristics between the extremes of the Arp\,220 and Mrk\,231
templates, although the dispersion in Fig.~1 must be due partially to the
spread in redshift. Over half of the sample display an Arp\,220-like upturn in
the mid-IR --- with a flat SED out to 8\,$\mu$m in the observed frame and a
sharp rise thereafter --- showing that active galaxies, although present, do
not dominate the MAMBO population. This is consistent with conclusions drawn
from UV/optical spectra and X-ray imaging \citep{cha03,ale03}.  Using
the model/empirical tracks to classify the eight MAMBO galaxies with measured
3.6--24-$\mu$m SEDs, Fig.~3 would indicate six starbursts and two AGN.  Thus
$\rm 75\pm 18$\% of the SMGs in our small sample have mid-IR colors
characteristic of high-redshift obscured starbursts. Their
mid-IR colors suggest a median redshift of $\sim$1.4 for our MAMBO sample,
somewhat lower than the $z=\rm 2.5$ measured for the radio-detected SMGs of
Chapman et al.\ (2003).  This may reflect differences in the selection
functions at 850 and 1200\,$\mu$m, with the latter selecting a somewhat
colder, lower redshift and intrinsically lower-luminosity section of the
population, or it may simply be the result of adopting inappropriate
SED templates.

If we assume the MAMBO sources lie at similar redshifts to the SCUBA
population and place the MAMBO sources at $z\sim 2.5$ then 8.0\,$\mu$m
corresponds to rest-frame $K$. It is therefore instructive to compare
the MAMBO galaxies with the $K$-corrected spectrum of Arp\,220
(Fig.~3). Arp\,220 has an absolute $K'$ magnitude of $-$24.4, roughly
1.3\,L${^\star}$ \citep{kim02}, and would have an 8.0-$\mu$m flux of
$\sim$2\,$\mu$Jy at $z=2.5$. The median 8.0-$\mu$m flux of the
$\ge$4$\sigma$ MAMBO sources is $\sim$20\,$\mu$Jy, implying naively
that these galaxies are already $\sim$15\,L$^{\star}$. The more
plausible alternatives are that these galaxies lie at lower redshifts
than we have assumed ($z\sim 1$--2, as implied by Fig.~3) or that their
observed 8.0-$\mu$m luminosity is dominated by massive, young stars or
hot dust.  Determining the exact cause of the apparently immense mid-IR
luminosities of these SMGs will require spectoscopic observations in
the optical and mid-IR to confirm their redshifts and the details of
their mid-IR spectral properties.

\section{Concluding remarks}

{\em Spitzer} observations between 3.6 and 24\,$\mu$m of nine galaxies
selected at 1200\,$\mu$m have improved our confidence in the
identification of secure counterparts, complementing what was learned
from radio imaging. Robust identifications have been possible for
$\sim$90\% of the MAMBO galaxies using 24-$\mu$m imaging, a similar fraction
to that achievable from radio imaging. Together, radio and mid-IR
imaging have yielded plausible counterparts for the entire MAMBO sample
in this region.  We conclude that 3.6--24-$\mu$m observations provide a
useful tool to aid in identifying SMGs.

Accretion onto an obscured central engine is sometimes betrayed by the shape of
the mid-IR continuum, confirming {\em Spitzer}'s potential to weed out active
galaxies. The fraction of mm galaxies with energetically-important AGN in our
small sample, 25\% (or 5--60\% at the 95\% confidence level),
is similar to that estimated from UV/optical spectroscopy and deep X-ray
imaging \citep{cha03,ale03}, but the majority of MAMBO galaxies have
rest-frame mid-/far-IR SEDs commensurate with obscured starbursts.

\acknowledgments This work is based in part on observations made with
the {\em Spitzer Space Telescope}, which is operated by the Jet
Propulsion Laboratory (JPL), California Institute of Technology
(Caltech) under NASA contract 1407. Support for this work was provided
by NASA through Contract Number \#960785 issued by JPL/Caltech.

%
%
%

%
% Table 1
%
\begin{deluxetable}{lccccccccl}
\rotate
\tabletypesize{\tiny}
\tablecaption{Catalog of MAMBO galaxies in the {\em Spitzer} Early
  Release Observations region}
\tablewidth{0pt}
\tablehead{
\colhead{Source\tablenotemark{a}}&
\colhead{$S_{1200} (\rm mJy)$}&
\colhead{$S_{850} (\rm mJy)$\tablenotemark{b}}&
\colhead{$S_{24} (\rm \mu Jy)$}&
\colhead{$S_{8.0} (\rm \mu Jy)$}&
\colhead{$S_{5.8} (\rm \mu Jy)$}&
\colhead{$S_{4.5} (\rm \mu Jy)$}&
\colhead{S$_{3.6} (\rm \mu Jy)$}&
\colhead{S$_{\rm 1.4GHz} (\rm \mu Jy)$}&
\colhead{Comment}
}
\startdata
{\em MM\,J105204.1+572658}&3.6$\pm$0.6&9.5$\pm$2.8&
249$\pm$31&18.6$\pm$1.9&27.6$\pm$3.3&20.2$\pm$2.1&15.2$\pm$1.6&36$\pm$5&(a); LE\,850.14\\
&&&---&18.5$\pm$1.9&22.2$\pm$3.0&14.7$\pm$1.5&9.8$\pm$1.0&72$\pm$8&(b)\\
{\em MM\,J105201.3+572448} & 3.4$\pm$0.6 &10.5$\pm$1.6&
193$\pm$31&10.2$\pm$1.7&11.8$\pm$2.8&8.2$\pm$0.9&3.8$\pm$0.8&73$\pm$8&LE\,850.01\\
{\em MM\,J105155.4+572310} & 3.3$\pm$0.8 &4.5$\pm$1.3&
125$\pm$32&26.4$\pm$2.7&19.9$\pm$2.8&5.8$\pm$0.8&6.1$\pm$0.8&47$\pm$5&LE\,850.18\\
{\em MM\,J105200.2+572425} & 2.4$\pm$0.6 &5.1$\pm$1.3&
534$\pm$55&112$\pm$12&79.0$\pm$8.0&72.6$\pm$7.3&72.0$\pm$7.2&$<$15&(a); $z=\rm 0.97$, LE\,850.08\\
&&&282$\pm$30&89.5$\pm$9.0&37.2$\pm$3.8&22.3$\pm$2.3&17.1$\pm$1.8&80$\pm$8&(b)\\
&&&162$\pm$30&17.8$\pm$1.8&17.2$\pm$2.8&20.7$\pm$2.1&26.7$\pm$2.7&22$\pm$6&(c)\\
\noalign{\smallskip}
\hline
\noalign{\smallskip}
{\bf MM\,J105157.6+572800}&2.2$\pm$0.6&$<$11&
121$\pm$30&24.5$\pm$2.5&35.7$\pm$3.8&29.8$\pm$3.0&44.5$\pm$4.5&21$\pm$6&\\
{\em MM\,J105207.2+572558}&1.7$\pm$0.5 &$<$12&
$<$90&15.0$\pm$1.7&14.5$\pm$2.8&17.3$\pm$1.8&12.8$\pm$1.3&23$\pm$6&\\
{\bf MM\,J105203.6+572612}\tablenotemark{c}& 1.7$\pm$0.5&$<$10&
166$\pm$30&12.1$\pm$1.7&15.8$\pm$2.8&12.0$\pm$1.2&7.9$\pm$0.8&19$\pm$6&\\
{\em MM\,J105216.0+572506} & 1.6$\pm$0.5&6.7$\pm$2.1&
578$\pm$59&---&---&---&---&48$\pm$5&LE\,850.29\\
{\em MM\,J105148.5+572408} & 2.1$\pm$0.7&$<$7.8&
211$\pm$30&16.7$\pm$1.7&16.8$\pm$2.8&13.4$\pm$1.4&11.9$\pm$1.2&29$\pm$6&\\
\enddata

\tablenotetext{a}{Italics signify sources for which robust counterparts have
been identified via radio imaging; bold type indicates a robust identification made
possible by the mid-IR data.}
\tablenotetext{b}{We quote either a detection or
a 3$\sigma$ limit based on the 850-$\mu$m noise measured in a
2.6-square-arcmin region centered on the mm source.}
\tablenotetext{c}{The source lies between LE\,850.04 and
LE\,850.14; in the SCUBA map it would thus have been affected by the off-beams
of both, hence its lack of detection at 850\,$\mu$m.}

\end{deluxetable}

\clearpage

\begin{figure}
\plotone{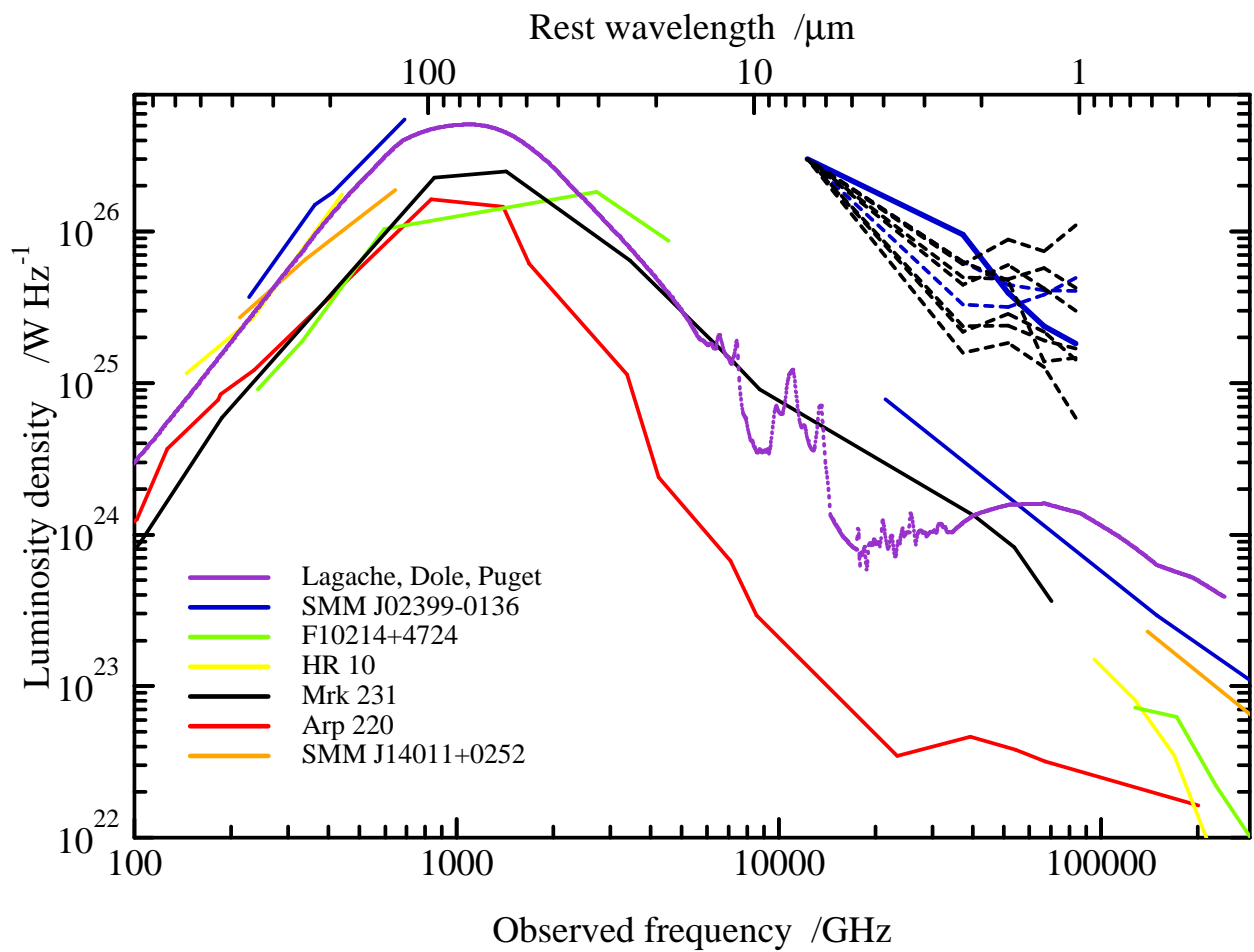}
\caption{Rest-frame SEDs of well-known dusty, luminous galaxies (see
Ivison et al.\ 2000) and the 10$^{13}$-L$_{\odot}$
synthetic starburst SED from \citet{lag03}.  Mid-IR SEDs for the 
MAMBO galaxies in our sample are shown as dashed lines (vertically
offset for clarity), normalised to give identical
24-$\mu$m luminosity densities and arbitrarily assuming a source
redshift of $z=\rm 2.5$.  We distinguish the three components of
MM\,J105200.2+572421.9 (a, b, c) by plotting them in blue, with a solid
line representing the power-law active component, `b'. SINGLE COLUMN.}
\label{fig1}
\end{figure}

\clearpage 
\begin{figure}
\plotone{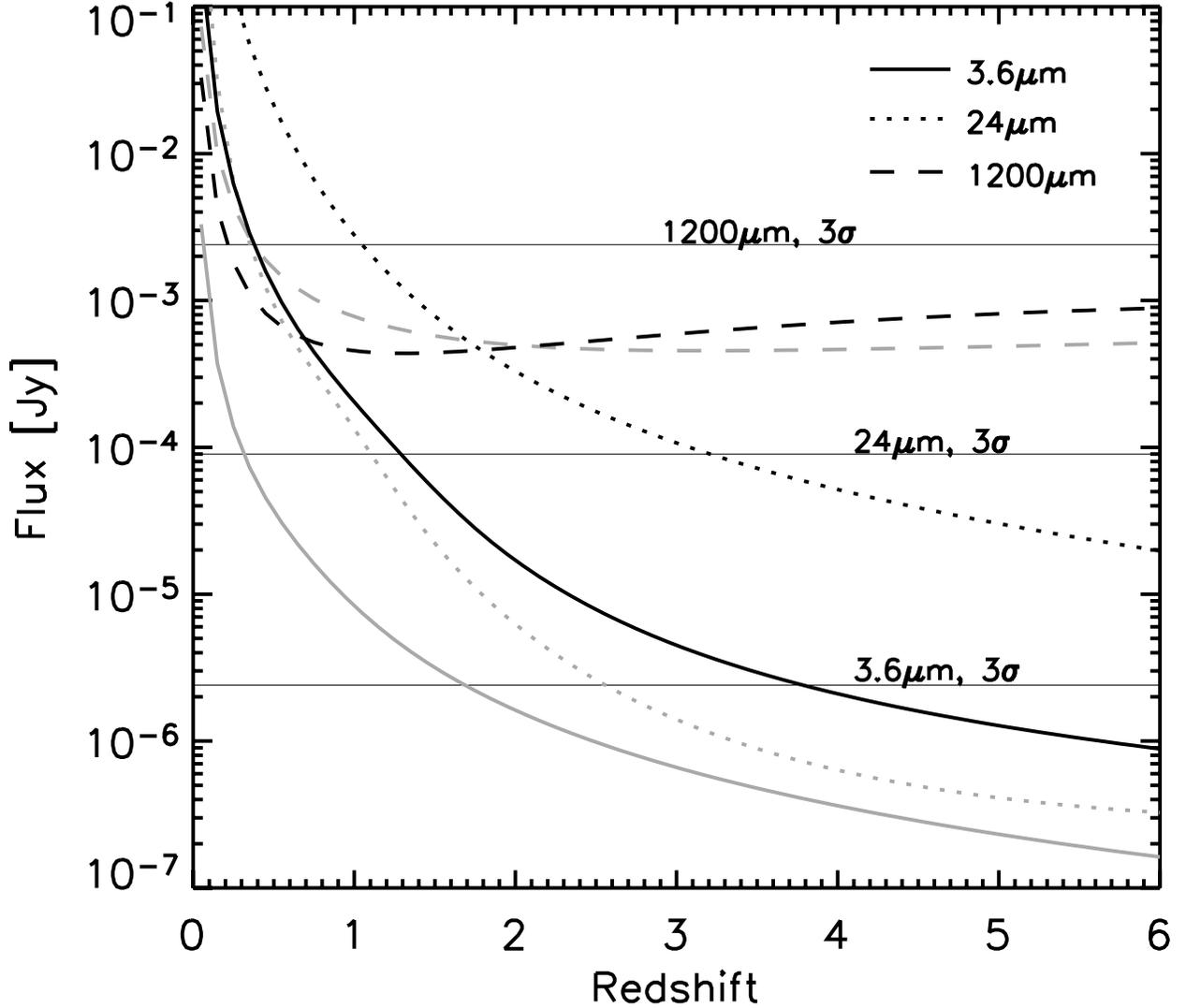}
\caption[]{The variation of flux density with redshift for Arp\,220
(pale lines) and the active galaxy, Mrk\,231 (dark lines). Three bands
are shown: 3.6, 24 and 1200\,$\mu$m, together with the 3$\sigma$
detection thresholds for our mid-IR/mm observations.  As can be seen,
our mid-IR observations are sufficiently sensitive to identify galaxies
with luminosities and SEDs similar to either Arp\,220 or Mrk\,231 out
to $z\gs\rm 1$--3.  We assume a cosmology with $\Omega_m=0.3$,
$\Omega_\Lambda=0.7$ and $H_o=70$\,km\,s$^{-1}$\,Mpc$^{-1}$. SINGLE
COLUMN.}
\label{fig2}
\end{figure} 

\clearpage 

\begin{figure}
\plotone{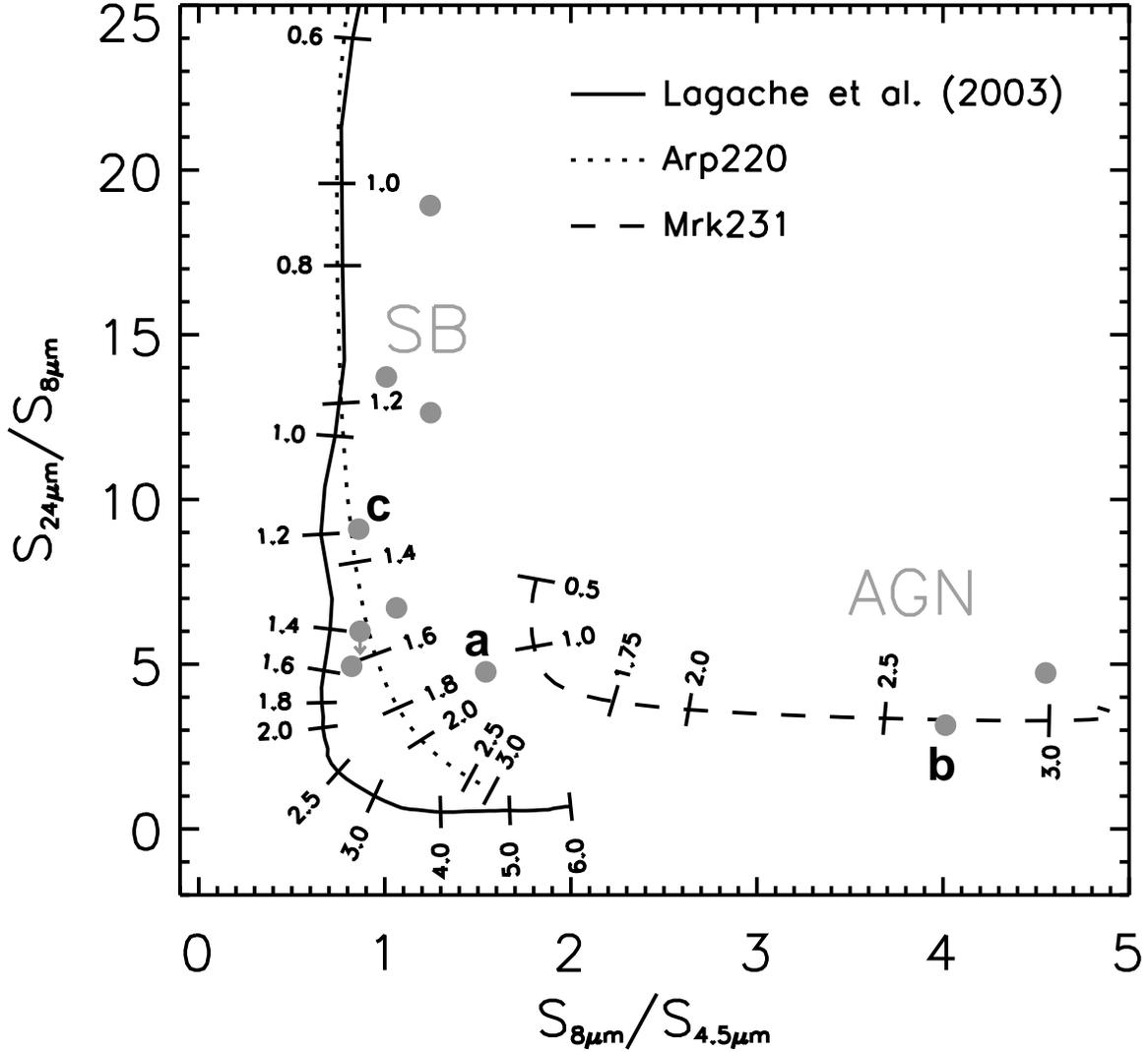}
\caption[]{A diagnostic color-color diagram showing $S_{24\mu \rm
m}/S_{8\mu \rm m}$ versus $S_{8\mu \rm m}/S_{4.5\mu \rm m}$. Tracks for
Arp\,220 (dotted line), a 10$^{13}$-L$_{\odot}$ synthetic starburst
(solid line), and Mrk\,231 (an active galaxy, dashed
line), illustrate how these ratios vary as they are moved from
$z\sim\rm 0$ to $z\gs\rm 3$. The diagnostic power of this plot lies in
the change in slope of the SED at rest-frame $\sim$3--4\,$\mu$m
observed in the spectrum of starbursts. For a submm/mm galaxy at
$z\sim\rm 2.5$, this results in a pivot around the 8-$\mu$m IRAC
waveband. Due to their power-law mid-IR continua, AGN at $z\ge\rm 1$
typically show higher $S_{8\mu \rm m}/S_{4.5\mu \rm m}$ at fixed
$S_{24\mu \rm m}/S_{8\mu \rm m}$ than starbursts, providing a simple
test of the power source.  We chose 4.5\,$\mu$m for the
short-wavelength band because it `comes free' with 8-$\mu$m IRAC data;
3.6\,$\mu$m would work equally well.  Mid-IR data for the nine detected
MAMBO sources are plotted as points, with the three components of
MM\,J105200.2+572421.9 plotted separately (identified as a, b, c).
These illustrate how the active galaxies are separated from the
starbursts. In theory, some degree of redshift discrimination is also
possible via this diagram (see \S4).
SINGLE COLUMN.}
\label{fig3}
\end{figure} 

\clearpage 

\begin{figure}
\plottwo{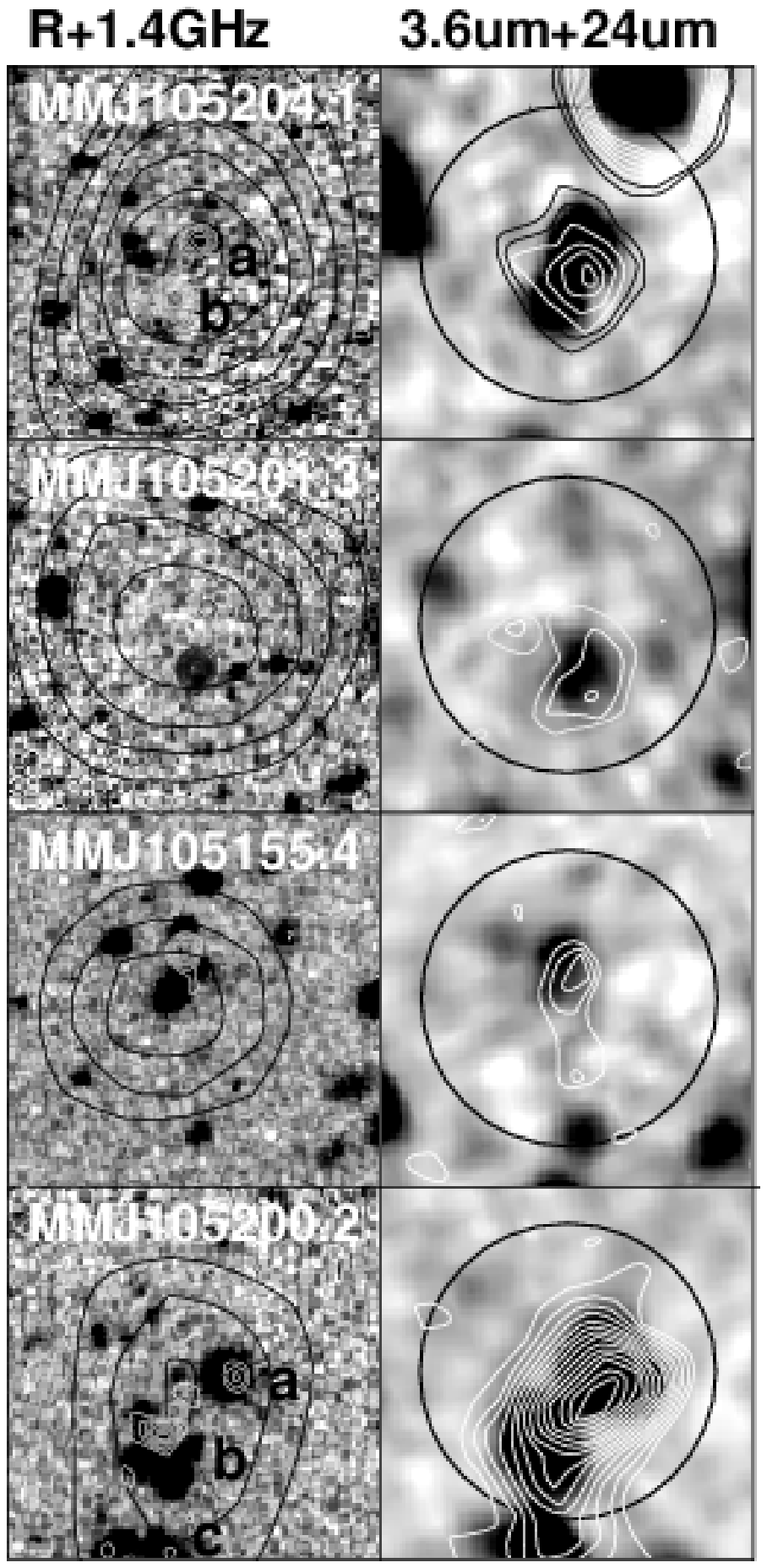}{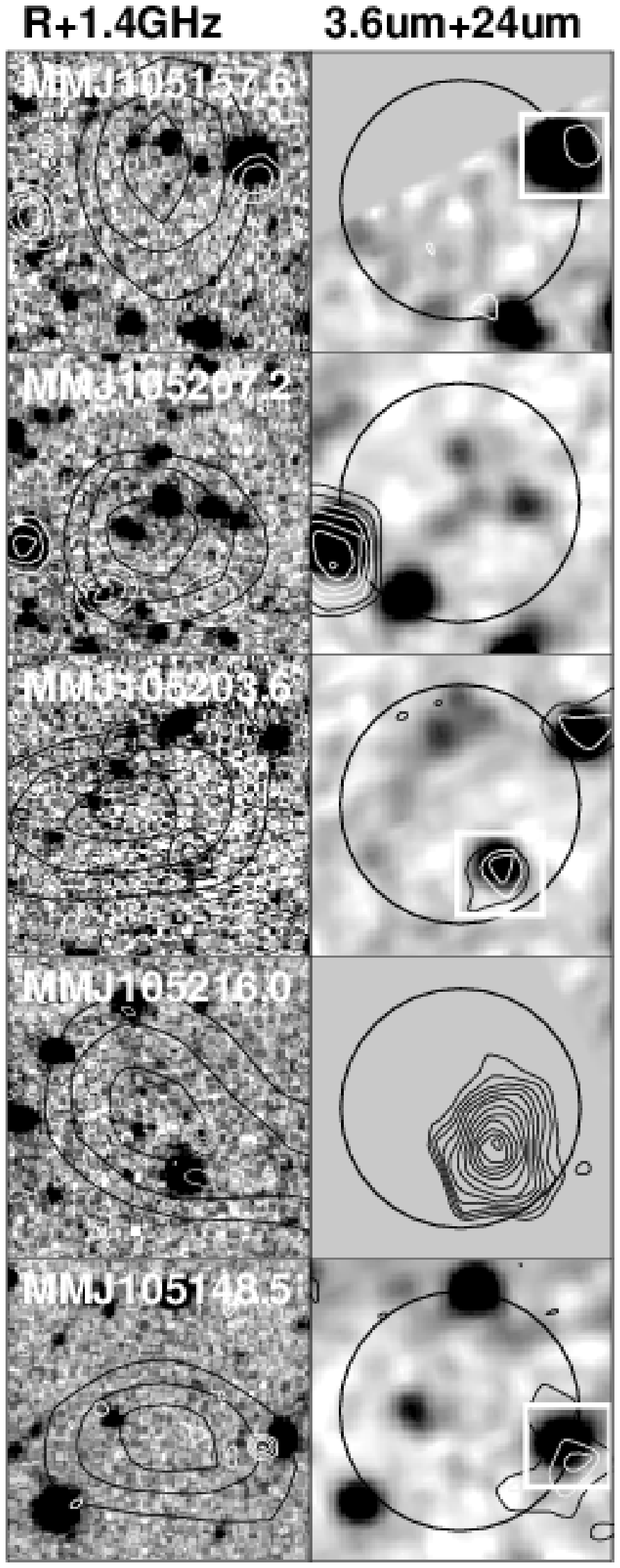}
\caption[]{Multi-wavelength images of $\rm 20'' \times 20''$ regions
centered on the nine MAMBO sources within the {\em Spitzer} field. For
each source we show: left-hand panel, $R$-band gray-scale  with
1200-$\mu$m contours overlaid at arbitrary levels and, at finer
resolution, 1.4-GHz contours at 2, 3, 4, 5, 6 $\times \sigma$, where
$\sigma = 5\,\mu$Jy beam$^{-1}$; right-hand panel, 3.6-$\mu$m
gray-scale with 24-$\mu$m contours (at 3, 4, 5, 6, 7 $\times \sigma$),
where the 3.6 and 24-$\mu$m images have been aligned. MAMBO sources are
marked by 8$''$-diameter circles in the right-hand panels. White
squares mark tentative identifications based on weak radio emission
near-coincident with 24-$\mu$m emission. FULL-PAGE PLATE.}
\label{fig4}
\end{figure}

\end{document}